# Simulation of Enhancement Mode GaN HEMTs with Threshold > 5 V using P-type Buffer

Sanyam Bajaj, Fatih Akyol, Sriram Krishnamoorthy, Ting-Hsiang Hung and Siddharth Rajan, *Member, IEEE*

*Abstract*—A high threshold voltage enhancement-mode GaN HEMT with p-type doped buffer is discussed and simulated. Analytical expressions are derived to explain the role of buffer capacitance in designing and enhancing threshold voltage. Simulations of the proposed device with p-type buffer show threshold voltages above 5 V, and a positive shift in threshold voltage as the oxide capacitance is reduced, thus enabling threshold voltage tunability over an unprecedented range for GaN-based HEMTs. The electric field profiles, breakdown performance, on-resistance and delay tradeoffs in the proposed *pGaN back HEMT* device are also discussed.

## I. Introduction

GROUP III-nitride based high electron mobility transistors (HEMTs) are suitable candidates for high-power electronics applications due to their high breakdown field, high-density and high-mobility two-dimensional electron gas (2DEG) [1,2]. These devices are inherently normally-on (depletion-mode or D-mode) due to the polarization in AlGaN, though normally-off (enhancement-mode or E-mode) operation has been researched actively with threshold voltages ($V_T$) up to ~2.5 V as shown in Figure 1 [3-10]. In this report we analyze the limitation of the conventional normally-off III-nitride HEMTs and propose a new approach for designing and achieving $V_T$ values higher than 5 V.

### A. Operation Principle of "pGaN back HEMT"

As known from the large body of work in Si metal-oxide-semiconductor field-effect-transistor (MOSFET) technology, $V_T$ is a function of oxide capacitance ($C_{OX}$) and channel doping. For example p-type doping ($N_A$) of the buffer an n-channel MOSFET affects the threshold voltage according to the well-known expression [11]

$$V_T = V_{FB} + \frac{2\varphi_{Fb}}{q} + \frac{\sqrt{2\varepsilon_S q N_A 2\varphi_{Fb}}}{C_{OX}}, \quad (1)$$

where $V_{FB}$ is the flat-band voltage, $\varphi_{Fb}$ is the difference between Fermi-energy ($E_F$) and intrinsic energy ($E_i$) in the bulk, and $\varepsilon_S$ is the semiconductor permittivity. The current approaches to achieve normally-off operation engineer the barrier layer, such as gate-recess [4-8], gate-injection transistor (GIT) using p-type layer [3,9] or ion-implantation under the gate [10]. However, the threshold voltage has no dependence on oxide or cap capacitance since the buffer layers are undoped in conventional HEMT structures, making the right-most term in equation (1) negligible. Therefore the threshold voltage is limited by the Fermi level position, $\varphi_{Fb}$ in the GaN buffer. This explains the large body of work in the literature where the threshold voltage in GaN HEMTs is limited to ~2.5 V, which corresponds to a semi-insulating GaN buffer with Fermi level pinned ~1 V above the valence band [12]. Using p-type GaN buffers, however, could provide a way to increase the threshold voltage beyond this limit.

To derive the analytical expressions for threshold voltage in HEMTs, we begin with Figure 2 which shows structure schematic of *pGaN back HEMT* and the associated charge, field and the energy-band profiles. For a normally-off device at equilibrium, the total applied gate-bias (threshold voltage),

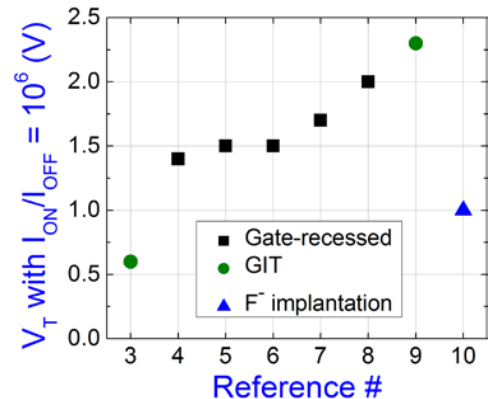

Fig. 1: A comparison of some of the highest reported threshold voltage ($V_T$) values in GaN HEMTs with undoped buffer, inferred from the reported semi-logarithmic $I_D$-$V_G$ transfer plots such that the $I_{ON}/I_{OFF}$ ratio is ~$10^6$.

Manuscript received January 22, 2015. This work was supported by ONR DEFINE MURI (N-00014-10-1-0937, Program Manager Dr. Paul A. Maki).

S. Bajaj, F. Akyol, S. Krishnamoorthy, T.-H. Hung and S. Rajan are with the Electrical and Computer Engineering Department, The Ohio State University, Columbus, OH 43210 USA (e-mail: bajaj.10@osu.edu).

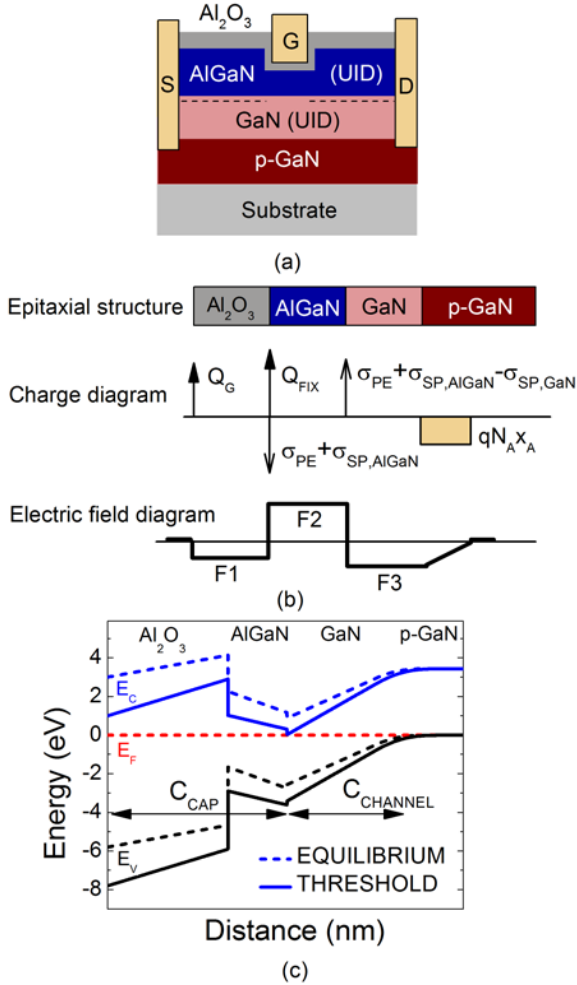

Fig. 2: *pGaN back HEMT* (Al$_2$O$_3$/AlGaN/GaN/p-GaN): (a) structure schematic; (b) associated charge and electric field diagrams; and (c) energy-band diagram under the gate at equilibrium and threshold conditions.

analytical expressions can be derived from the energy-band diagram of Al$_2$O$_3$/AlGaN/GaN/p-GaN structure shown in Figure 2(c) as

$$V_T - \frac{\phi_b}{q} = F_1 t_1 - \frac{\Delta E_{C1}}{q} + F_2 t_2 - \frac{\Delta E_{C2}}{q}, \qquad (3)$$

$$F_3 t_3 + \frac{N_A x_A^2}{2\varepsilon_S} = \frac{E_g}{q} - (\frac{E_F}{q} - \frac{E_V}{q}) \qquad (4)$$

Here $\phi_b$ is the schottky barrier height, $\Delta E_{C1}$ is the Al$_2$O$_3$/AlGaN conduction band offset, $\Delta E_{C2}$ is the AlGaN/GaN conduction band offset, $x_A$ is the depletion width in the p-type buffer, $E_g$ is the bandgap of GaN, $E_V$ represents the energy at the valence band edge, F$_1$ is the electric field in the first cap layer (Al$_2$O$_3$) of thickness t$_1$, similarly F$_2$ and t$_2$ for the second cap layer (AlGaN), and F$_3$ and t$_3$ for the third (GaN channel) layer. From charge and field diagrams shown in Figure 2(b), the expressions for electric fields in the layers can be derived from Poisson's equation as

$$F_1 = F_2 + \frac{(\sigma_{PE} + \sigma_{SP,AlGaN})/q - Q_{FIX}}{\varepsilon_S}, \qquad (5)$$

$$F_2 = F_3 - \frac{\sigma_{PE} + \sigma_{SP,AlGaN} - \sigma_{SP,GaN}}{q\varepsilon_S}, \qquad (6)$$

$$F_3 = \frac{qN_A x_A}{\varepsilon_S} \qquad (7)$$

Here $\sigma_{PE}$ is the piezoelectric polarization charge at the AlGaN/GaN interface, $\sigma_{SP,A}$ is the spontaneous polarization charge of AlGaN, $Q_{FIX}$ is the fixed positive sheet charge at the Al$_2$O$_3$/AlGaN interface [6], $q$ is the electron charge of 1.6x10$^{-19}$ C, and $\sigma_{SP,G}$ is the spontaneous polarization charge of GaN. These equations provide a route to achieving higher threshold voltage in Al$_2$O$_3$/AlGaN/UID GaN/p-GaN type metal-insulator-semiconductor (MIS) HEMT structures. Choosing the appropriate cap and channel capacitances can enable us to achieve threshold that is much higher than it is possible in undoped channel HEMTs. The remainder of this work explores *pGaN back HEMT* in detail using 2D device simulations.

as seen in Figure 2(c), drops across the cap layers (V$_{CAP}$) and the buffer, or the channel layer (V$_{CHANNEL}$).

$$V_T = V_{CAP} + V_{CHANNEL} = V_{CHANNEL}(\frac{C_{CHANNEL}}{C_{CAP}} + 1), \qquad (2)$$

where C$_{CAP}$ and C$_{CHANNEL}$ are the capacitances of the cap layers and the channel layers respectively. Since the undoped buffer in standard HEMTs is fully depleted, it offers negligible capacitance (C$_{CHANNEL}$ << C$_{CAP}$), implying that the entire V$_T$ drops across the buffer (V$_T$ = V$_{CHANNEL}$). Replacing undoped buffer with a p-type doped one offers larger capacitance C$_{CHANNEL}$, thereby enhancing V$_T$ by the factor of $\frac{C_{CHANNEL}}{C_{CAP}} + 1$. In other words, p-type buffer offers large C$_{CHANNEL}$, and enables the cap layers (C$_{CAP}$) to design and tune the threshold voltage of the device, for instance increasing cap thickness would reduce C$_{CAP}$ and increase threshold voltage, V$_T$. Assuming device threshold condition to be when the channel conduction-band touches the Fermi-level, the

## II. SIMULATION

### A. Device Structures and Models

Simulations for the *pGaN back HEMT* were carried out using the 2-D device simulator Silvaco ATLAS [13]. The gate-recessed structures consisted of Al$_2$O$_3$ as gate-dielectric with varying thicknesses (10, 20, 30 and 40 nm), 40 nm of Al$_{0.15}$Ga$_{0.85}$N cap, recessed down to 5 nm under the gate, followed by 15 nm of UID GaN channel (n-type dopant concentration $10^{15}$ cm$^{-3}$), 200 nm of p-doped GaN buffer layer (p-type dopant concentration $5 \times 10^{18}$ cm$^{-3}$) and 1 μm insulating GaN substrate. The gate-length, gate-source and the gate-drain spacings were fixed at 1 μm, 1 μm and 4 μm respectively. We used Ni/Al$_2$O$_3$ barrier height of 3 eV [14], and a fixed positive sheet charge density (Q$_{FIX}$) of $1.1 \times 10^{13}$ cm$^{-2}$ at the Al$_2$O$_3$/AlGaN interface for the simulations [6]. The composition-dependent material structure, spontaneous polarization, and piezoelectric polarization parameters were included in the simulation in order to introduce the correct polarization charges at the hetero-interfaces [15]. The models that were used in the simulations include Shockley-Read-Hall model (SRH), parallel electric field model for velocity saturation (FLDMOB), Fermi-Dirac model (FERMI), polarization model (SPONTANEOUS) and incomplete ionization model (INCOMPLETE), while the Farahmand Modified Caughey-Thomas expression for group-III nitride material system [16,17] was applied with a saturation velocity of $10^7$ cm/s, and a channel hall mobility of 1500 cm$^2$/Vs which can be considered as a typical value in GaN HEMTs. In reality, the channel mobility in the device may be lower due to remote ionized impurity scattering, more on which has been discussed in later sections. The p-type doped buffer was simulated with incomplete ionization model due to the nature of the dopant material, usually Magnesium in the case of III-nitrides. Finally, body contact was used to ground the p-type buffer, while all "undoped" layers in the simulator were doped with a low donor background concentration of $10^{15}$ cm$^{-3}$.

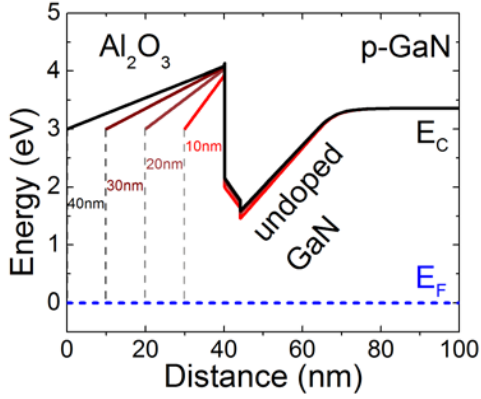

Fig. 3: Conduction band profiles (under the gate) for Al$_2$O$_3$/AlGaN/UID GaN/p-GaN MIS-HEMT structure with varying Al$_2$O$_3$ thicknesses of 10, 20, 30, and 40 nm, at equilibrium condition.

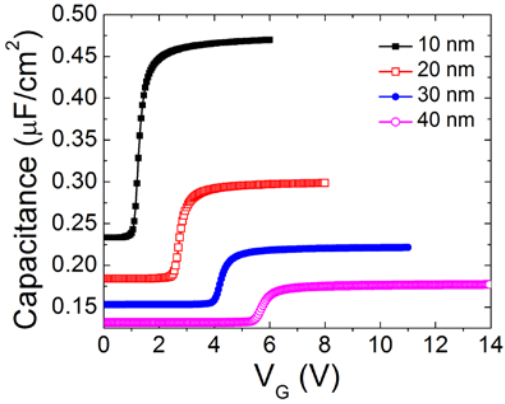

Fig. 4: Simulated quasi-static capacitance-voltage characteristics with varying Al$_2$O$_3$ thicknesses of 10, 20, 30, and 40 nm, illustrating 2DEG accumulation and a positive shift in threshold as the oxide thickness is increased.

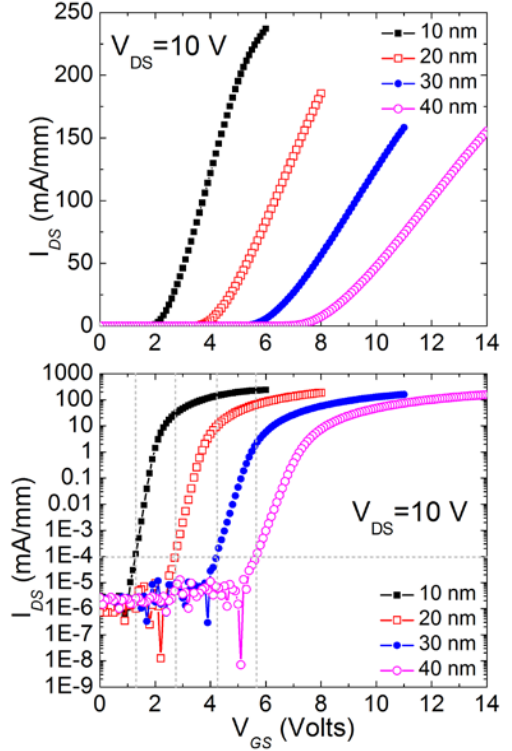

Fig. 5: Simulated I$_D$-V$_G$ transfer characteristics with varying Al$_2$O$_3$ thicknesses of 10, 20, 30, and 40 nm in linear and semi-logarithmic scales, suggesting a positive shift in the threshold voltage with increasing cap thickness or decreasing capacitance.

## III. RESULTS AND DISCUSSIONS

### A. Threshold Voltage

Figure 3 shows the equilibrium conduction band profiles of the simulated Al$_2$O$_3$/AlGaN/UID GaN/p-GaN MIS-HEMT structures with varying Al$_2$O$_3$ thicknesses of 10, 20, 30 and 40 nm. Figure 4 shows the simulated quasi-static capacitance-voltage characteristics for the four cases. The maximum positive gate-bias was varied for each case such that the field in the oxide layer is approximately 3 MV/cm, i.e. below the critical field for the onset of Fowler-Nordheim tunneling [18]. The depletion capacitance in the C-V curves (Figure 4) corresponds to the entire depletion width in the device, i.e. C$_{CAP}$ in series addition with C$_{CHANNEL}$, while the accumulation capacitance in the curves corresponds to the cap capacitance, i.e. $C_{CAP}$ only. The curves also show a positive shift in the threshold voltage (onset of accumulation) as the oxide thickness is increased, in addition to the obvious reduction in the accumulation capacitance $C_{CAP}$. Figure 5 shows the simulated transfer characteristics with varying oxide thickness in both linear and semi-logarithmic scale. The transfer curves show a similar positive shift in the threshold voltage as the oxide thickness is increased, whereas the lower capacitance, $C_{CAP}$, which translates to a lower transconductance is also evident from the slope of the curves. This reduction in the transconductance is the direct tradeoff that results from the operation of pGaN back HEMT. The threshold voltage values extracted ($I_{ON}/I_{OFF}$ ratio of ~$10^6$) from the semi-logarithmic $I_D$-$V_G$ curves shown in Figure 5 were 1.2, 2.8, 4.2 and 5.7 V for $Al_2O_3$ thicknesses of 10, 20, 30 and 40 nm respectively. These values are higher than any experimental or simulated GaN HEMT report, and suggest that p-type buffers could be a highly promising approach to achieving large threshold voltages GaN HEMTs.

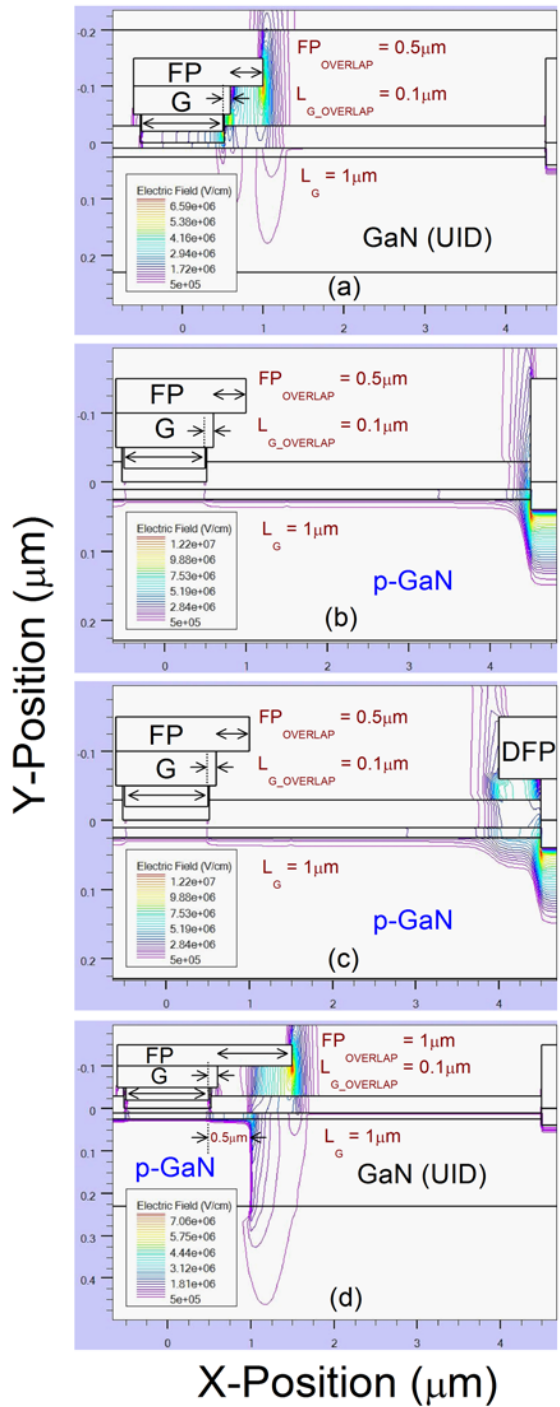

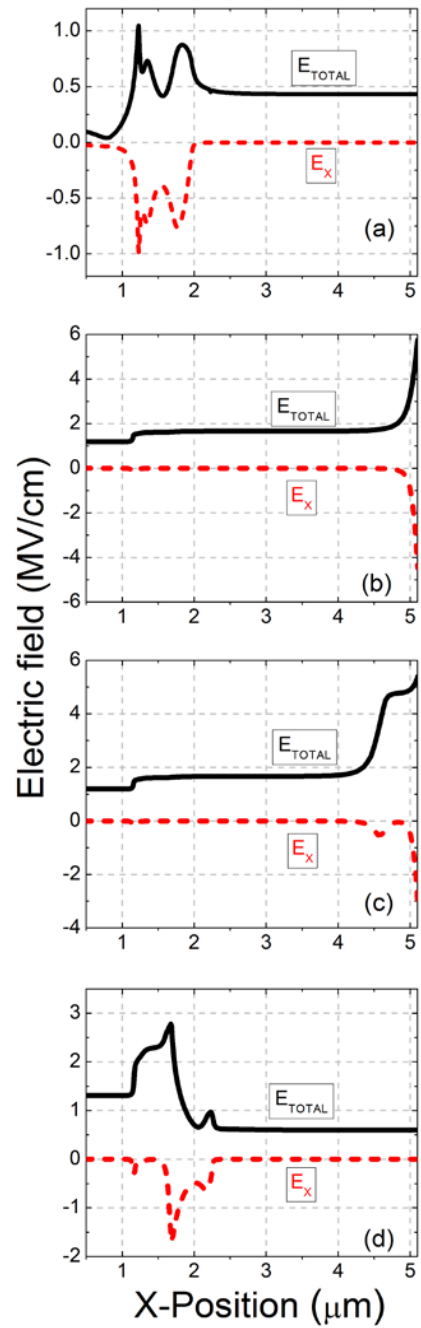

Fig. 6: Contour maps of the distribution of electric field in simulated HEMT structures with (a) conventional undoped buffer; (b) p-type doped buffer; (c) p-type doped buffer and a drain-connected field plate; and (d) p-type doped buffer only under the source and gate. In all cases, a high drain bias was applied ($V_{DS}$ = 50 V) at off-state condition ($V_{GS}$ = 0 V) to understand 3-terminal breakdown characteristics. (FP: gate field plate; DFP: drain field plate)

Fig. 7: Electric field profile, total ($E_{TOTAL}$) and x-component ($E_X$), across the channels of the simulated HEMTs corresponding to the four cases (a) to (d) in figure 6. In all cases the cutline was taken 1 nm below the AlGaN/GaN interface.

### B. On-resistance

The operation of pGaN back HEMT, which utilizes p-type buffer, raises the conduction band profile and depletes the 2DEG density. Since the intrinsic on-resistance of the device is inversely proportional to the product of 2DEG charge density and mobility ($n_s*\mu$), it is another critical tradeoff offered by the proposed device. To enhance the charge density in access regions, a higher composition AlGaN barrier, a thicker barrier or modulation doping could be utilized. This can be achieved by designing the device growth with a thinner and lower composition barrier (needed for the intrinsic region) followed by a higher composition AlGaN barrier with additional modulation doping which can be recessed under the gate. This may also be achieved by the complex yet realizable regrowth techniques. The channel mobility in enhancement-mode GaN MISHEMTs is limited in the intrinsic region of the device by phonon scattering and remote ionized impurity scattering. In addition, the channel of the proposed device will also experience remote scattering from the ionized dopant concentration in the buffer. This effect, however, could be eliminated using an undoped spacer layer thicker than approximately 5 nm [19] between the channel and p-type buffer, like the design used in this work. Furthermore, the high field in the undoped GaN channel, as seen in Figure 3, could also degrade the mobility as it pushes the 2DEG wavefunction towards the barrier layer and magnifies the effect of interface roughness. This, however, could be mitigated by adding AlN binary interlayer above the GaN channel [20], or by having a thicker undoped channel layer to reduce this field. Thus, even though the operation of the proposed device suggests higher on-resistance as a tradeoff, an optimized design could well result in on-resistance values comparable to the existing state-of-the-art normally-off GaN MISHEMTs.

### C. Electric Field profile

The use of p-type buffer results in another critical tradeoff between the threshold voltage and the blocking voltage of the device. Figure 6 shows the electric field profiles (contour maps) of four unique gate-recessed HEMT structures at $V_{DS} =$ 50 V in the off-state ($V_{GS} = 0$ V) with 10 nm $Al_2O_3$ gate-dielectric. The first structure consisted of undoped buffer, the second one consisted of p-type doped buffer (dopant concentration of $5x10^{18}$ cm$^{-3}$), the third one consisted of p-type doped buffer (dopant concentration of $5x10^{18}$ cm$^{-3}$) and a drain-connected field plate, and the fourth one consisted of p-type buffer only under gate and source. Figure 7 shows the electric field profiles, total and lateral (x-component), across the channels corresponding to the four cases in Figure 6 respectively. Unlike conventional HEMTs, where the peak electric field occurs at the gate-edge towards the drain, p-type doped buffer causes the peak electric field at the reverse biased drain-edge as seen in Figures 7(a) and (b). The magnitude in the latter case is comparatively higher and may be reduced by lowering the acceptor concentration, although it would result in a lower threshold voltage as the aforementioned tradeoff. Adding a drain-connected field plate also helps to improve the field profile in the channel, as seen in Figure 7(c), although the effect is less significant with additional plates (not shown here). The final structure in Figure 7(d) consists of p-type buffer patterned only under source and drain in addition to a gate-connected field plate. It can be seen in Figure 5(d) that this structure further reduces the peak electric fields in the device, suggesting a higher breakdown voltage. Thus the *pGaN back HEMT* design in Figure 6(d) and 7(d) could improve the device breakdown performance in addition to achieving very high threshold voltages. This design can be realized and further improved using regrowth or fabrication techniques such as graded doping concentration under the gate, additional field plates or guard rings.

### D. Switching Delay

The intrinsic delay of the device is limited by the capacitive charging delay at the gate (RC delay). Since the hole mobility is much lower (about two orders of magnitude) than the electron mobility in GaN, the resistance of p-type buffer layer is much higher than the channel resistance and hence dominates. Assuming a bottom ohmic contact to the p-type buffer, the resistance of the p-type layer can be calculated as

$$R = \frac{\rho L_P}{A} = \frac{1}{q\mu_P P_0} * \frac{L_P}{A}, \qquad (8)$$

where $\rho$ is the resistivity, $L_P$ is the thickness of the buffer, $\mu_P$ is the hole mobility, $P_0$ is the hole concentration, and A is the device area. Using the total gate capacitance C, the delay is therefore estimated as

$$R*C = \frac{\varepsilon_S}{q\mu_P P_0} * \frac{L_P}{t}, \qquad (9)$$

where $t$ is the total thickness contributing to the gate capacitance. Assuming hole mobility of 10 cm$^2$/Vs and hole concentration of $10^{18}$ cm$^{-3}$, the ratio $L_P / t \sim 1$, the device delay is estimated to be in the order of 10 ps, which is longer than that of conventional HEMTs, nevertheless suitable for high power electronics and switching applications.

## IV. CONCLUSION

Enhancement-mode AlGaN/GaN HEMT with p-type doped buffer was investigated and simulated using 2D device simulations. The constraint of conventional undoped buffer and the role of buffer capacitance was discussed using analytical expressions of threshold voltage for $Al_2O_3$/AlGaN/GaN/p-GaN MIS-HEMT structures.

Simulations showed high threshold voltages up to 5.7 V, and a positive shift as the oxide capacitance is reduced, thus enabling its tunability over an unprecedented range for GaN-based HEMTs. The on-resistance of the device was analyzed and estimated to be comparable to conventional HEMTs. The electric field profiles suggested a lower breakdown in the case of doped buffer as compared to the case of undoped one, which may be overcome by optimizing the buffer design such as lowering the acceptor concentration, patterning it under the source and gate, and adding field plates. Finally, the RC delay in the *pGaN back HEMT* was calculated to estimate the switching performance.

These results demonstrate the potential of the proposed *pGaN back HEMT* in achieving threshold that is much larger than it is possible in conventional HEMT, and encourage further research for the next generation enhancement mode HEMTs for power-switching applications.